\documentclass[journal]{IEEEtran}
\usepackage{amssymb}
\usepackage{amsmath,amsfonts}
\usepackage{array}
\usepackage[caption=false,font=normalsize,labelfont=sf,textfont=sf]{subfig}
\usepackage{textcomp}
\usepackage{stfloats}
\usepackage{url}
\usepackage{verbatim}
\usepackage{graphicx}
\usepackage{cite}
\usepackage{enumitem}
\usepackage[linesnumbered,vlined,ruled,commentsnumbered]{algorithm2e}
\usepackage{algpseudocode}
\SetKwRepeat{Do}{do}{while}

\SetAlFnt{\small}
\SetAlCapFnt{\small}
\SetAlCapNameFnt{\small}
\IncMargin{-\parindent}
\SetKw{Break}{break}
\usepackage{amsmath}
\usepackage{booktabs} 
\usepackage{caption} 
\usepackage{multirow}
\usepackage{graphicx}
\usepackage{xcolor}
\usepackage{hyperref}
\hypersetup{
    colorlinks=true,
    linkcolor=blue
    }

\newcommand{\virgolette}[1]{``#1''}
\usepackage{pifont}

\begin{document}

\title{Dynamic Hashtag Recommendation in Social Media with \\Trend Shift Detection and Adaptation}

\author{Riccardo Cantini, Fabrizio Marozzo,~\IEEEmembership{Senior Member,~IEEE,}
Alessio Orsino, \\Domenico Talia,~\IEEEmembership{Senior Member,~IEEE,}
and Paolo Trunfio,~\IEEEmembership{Senior Member,~IEEE}
\thanks{Riccardo Cantini, Fabrizio Marozzo, Alessio Orsino, Domenico Talia, and Paolo Trunfio are with the DIMES Department, University of Calabria, Italy.
\\E-mail: \{rcantini, fmarozzo, aorsino, talia, trunfio\}@dimes.unical.it}
\thanks{Corresponding author: R. Cantini}
\thanks{Manuscript received April, 2025.}}

\markboth{
}
{Cantini \MakeLowercase{\textit{et al.}}: Dynamic hashtag recommendation in social media with trend shift detection and adaptation}


\maketitle

\IEEEpubid{\begin{minipage}{\textwidth}
    \vspace{1.5cm}
    \centering
    \fontsize{7.5}{8}\selectfont This work has been submitted to the IEEE for possible publication. Copyright may be transferred without notice, after which this version may no longer be accessible.
\end{minipage}}

\begin{abstract}
Hashtag recommendation systems have emerged as a key tool for automatically suggesting relevant hashtags and enhancing content categorization and search. However, existing static models struggle to adapt to the highly dynamic nature of social media conversations, where new hashtags constantly emerge and existing ones undergo semantic shifts. To address these challenges, this paper introduces H-ADAPTS (\textit{Hashtag recommendAtion by Detecting and adAPting to Trend Shifts}), a dynamic hashtag recommendation methodology that employs a trend-aware mechanism to detect shifts in hashtag usage---reflecting evolving trends and topics within social media conversations---and triggers efficient model adaptation based on a (small) set of recent posts. Additionally, the Apache Storm framework is leveraged to support scalable and fault-tolerant analysis of high-velocity social data, enabling the timely detection of trend shifts. Experimental results from two real-world case studies, including the COVID-19 pandemic and the 2020 US presidential election, demonstrate the effectiveness of H-ADAPTS in providing timely and relevant hashtag recommendations by adapting to emerging trends, significantly outperforming existing solutions.
\end{abstract}


\section{Introduction}  
\IEEEPARstart{T}{he} widespread use of social media has fostered global connections and generated vast amounts of data, offering key tools for analyzing user behavior and sentiment, while also posing the need for efficient categorization and search mechanisms. \cite{rangarajan2024social, belcastro2020learning, saxena2024depth}. A common tool for organizing content is the \textit{hashtag}---a string preceded by the \virgolette{\#} symbol---used to label posts and link them to trending topics or broader conversations, enhancing content discoverability and helping form communities around shared interests. However, the informal writing style typical of social media and the unrestricted nature of hashtag selection often make it difficult for users to choose relevant hashtags. This results in many posts lacking representative hashtags, which hampers effective categorization and retrieval. To mitigate this issue, hashtag recommendation systems have emerged to automatically suggest relevant hashtags, improving content relevance and user engagement \cite{yakovlev2024recommendation,HASHET-TKDD-2022}.

Effective hashtag recommendation on microblogging platforms requires adaptive, trend-aware systems due to the real-time, dynamic nature of user-generated content.
Adaptiveness refers to a model’s ability to adjust to shifting data distributions, an issue known as \textit{concept drift} \cite{yuan2022recent,su2024elastic}. In the context of hashtag recommendation, this enables the model to suggest contextually relevant hashtags that reflect the latest trends in online conversations.
While some work has explored adaptive models in domains like pictures and news article tagging \cite{kolyszko2024dynamic, hastagger+,gupta2025harnat}, real-time hashtag recommendation in microblogging remains underexplored. Static models trained on fixed datasets often struggle to handle emerging or semantically evolving hashtags, leading to degraded performance over time. However, designing adaptive models can be challenging due to the volume and velocity of social media data \cite{gupta2018unleashing, belcastro2022programming, amen2022big}. Approaches relying on regular retraining can be inefficient, as frequent retraining is computationally expensive and may still fail to adapt promptly to real-time shifts.  Moreover, many existing methods lack integration with big data frameworks, limiting their practical applicability in real-world scenarios.

To tackle these challenges, we propose \textit{H-ADAPTS} (\textit{Hashtag recommendAtion by Detecting and adAPting to Trend Shifts}), a dynamic hashtag recommendation method tailored to the ever-evolving nature of social media. H-ADAPTS uses real-time big data processing to detect shifts in trending hashtags and semantic changes in existing ones, enabling rapid and efficient adaptation. The underlying recommendation model is \textit{HASHET} (\textit{HAshtag recommendation using Sentence-to-Hashtag Embedding Translation}), a semi-supervised approach introduced in \cite{HASHET-TKDD-2022}. The key contributions of our work are:

\begin{itemize}
    \item A \textit{trend-aware} mechanism that detects real-time trend shifts using a variation of the Jaccard Distance to measure dissimilarity between ranked sets of top-$n$ hashtags. This allows the system to adapt only when meaningful shifts occur, avoiding unnecessary resource-intensive retraining.

    \item An efficient adaptation strategy that retrains the model on a (small) sliding window of recent posts jointly using transfer learning and progressive fine-tuning, ensuring smooth adaptation while minimizing computational cost.

    \item To manage high-velocity social data, H-ADAPTS employs Apache Storm for real-time, scalable, and fault-tolerant processing of unbounded data streams, enabling timely trend shift detection and relevant suggestions.

    \item We conducted extensive experiments on two real-world case studies, centered on the COVID-19 pandemic and 2020 US presidential election, analyzing detected trend shifts and demonstrating the model’s ability to effectively follow the dynamicity of the online conversation.
\end{itemize}

The remainder of the paper is organized as follows. Section \ref{sec:related_work} reviews related work. Section \ref{sec:h-adapts} provides a detailed description of H-ADAPTS. Section \ref{sec:deployment} outlines the Storm-based system design. Section \ref{sec:performance} presents the experimental evaluation. Finally, Section \ref{sec:conclusion} concludes the paper.

\subsection{Problem Formulation}

The hashtag recommendation task is aimed at learning a model $\mathcal{M}$ such that $\mathcal{M}(p) = \mathcal{T}^k_p \subseteq \mathcal{H}$, where $\mathcal{T}^k_p = \{t^1_p, t^2_p,\dots, t^k_p\}$ is the set of the $k$ target hashtags to be recommended for post $p \in \mathcal{P}$, while $\mathcal{P}$ and $\mathcal{H}$ are the sets of all possible posts and hashtags, respectively.

In dynamic environments like social media platforms, the emergence of new hashtags that link trending topics and events with a high social impact can affect the recommendation accuracy of model $\mathcal{M}$. Therefore, it must evolve over time to account for fluctuations in the relevance of hashtags chosen by social users to label published content. Here, this phenomenon is referred to as \textit{trend shift}, which we model as a form of concept drift, and it denotes a change in the hashtag relevance at two different time points, $t^{\prime}$ and $t^{\prime\prime}$.
We define the set of hashtags relevant at time $t$ as $\mathcal{H}^t = \{h \in \mathcal{H} \mid \psi^t_h > 0\}$, where $\psi^t_h$ is the relevance of a given hashtag $h \in \mathcal{H}$ at time $t$.
Given $\mathcal{H}^{\prime}$ and $\mathcal{H}^{\prime\prime}$, the sets of hashtags relevant at times $t^{\prime}$ and $t^{\prime\prime}$, respectively, a trend shift is detected when $\delta(\mathcal{H}^{\prime}, \mathcal{H}^{\prime\prime}) > \omega$, where $\delta$ is a distance metric between sets and $\omega$ is a predefined threshold. Therefore, a shift can be driven by two factors:
\begin{enumerate}
    \item A new hashtag $h_{\text{new}} \in \mathcal{H}^{\prime\prime}$ emerges such that $h_{\text{new}} \notin \mathcal{H}^{\prime}$ (i.e., $\psi^{t^{\prime}}_{h_{\text{new}}} = 0$).
    \item The relevance of an existing hashtag $h \in \mathcal{H}^{\prime} \cap \mathcal{H}^{\prime\prime}$ changes over time, i.e., $\lvert \psi^{t^{\prime}}_h - \psi^{t^{\prime\prime}}_h \rvert \,> 0$.
\end{enumerate}

Once a trend shift is detected, the recommendation model $\mathcal{M}$ must be adapted to reflect the new trending hashtags and their relevance, thereby maintaining its recommendation abilities. Adaptation can be performed by fine-tuning the model's parameters $\theta$ on a set of examples $\mathcal{D}_{t^{\prime\prime}}$ that incorporate the updated post-hashtag distribution for relevant hashtags, i.e., the pairs $(p, \mathcal{T}^k_p)$. The adapted model is therefore obtained by minimizing a suitable loss function $\ell$:

\begin{equation}
\min_{\theta} \mathbb{E}_{(p, \mathcal{T}^k_p) \sim \mathcal{D}_{t^{\prime\prime}}} [\ell(\mathcal{M}(p, \theta), \mathcal{T}^k_p)]
\end{equation}

\section{Related Work}
\label{sec:related_work}
Hashtag recommendation has gained significant attention in recent years due to the growing volume of user-generated content on social media. These systems aim to help users find and use relevant, popular hashtags to enhance post visibility, supporting content categorization and search \cite{yakovlev2024recommendation, chakrabarti2023hashtag, HASHET-TKDD-2022}. Here we review key contributions in the field, grouping state-of-the-art techniques into three main categories based on the followed approach.

\smallskip \noindent   
\textbf{Unsupervised Models.}
Unsupervised approaches to hashtag recommendation aim to extract meaningful features from unlabeled data to suggest relevant hashtags. For instance, the Hashtag Frequency Inverse Hashtag Ubiquity (HF-IHU) method \cite{otsuka2016hashtag} is a TF-IDF variation that uses hashtag ubiquity across the corpus to guide recommendations, leveraging Apache Hadoop for scalable Twitter stream processing. In \cite{ben-lhachemi2018}, tweets are represented as the average of their word embeddings and clustered using DBSCAN to identify semantically related groups. Hashtags are then recommended based on their proximity to the cluster centroids.
Probabilistic topic models are also commonly used to uncover latent topic distributions in posts and recommend relevant hashtags. For example, Godin et al. \cite{godin2013topic} leveraged Latent Dirichlet Allocation (LDA) \cite{blei2003latent} for content-based recommendation of general hashtags, treating documents as mixtures of topics.
Additionally, personalized unsupervised models employ Bayesian Personalized Ranking (BPR) to learn hashtag relevance from user features and past behavior. As an example, the Microtopic Recommendation Model (MTRM) \cite{Li2017personalized} introduces a probabilistic latent factor model that integrates user behavior, hashtags, content, and contextual information, combining collaborative filtering, content analysis, and feature regression.

\smallskip \noindent\textbf{Supervised Models.}
Among supervised models, many exploit attention mechanisms to generate semantically-rich representations by dynamically focusing on relevant parts of the input sequence, enabling contextualized hashtag recommendations.
For instance, Li et al. proposed the Topical Co-Attention Network (TCAN) \cite{li2019topical}, a neural model that jointly captures content and topical information. Similarly, tSAM-LSTM (Temporal enhanced sentence attention model-LSTM) \cite{ma2018temporal}, extends LSTM with sentence-level attention informed by the temporal dynamics of microblogs.
Besides attention-based models, other supervised techniques exist that rely on learning-to-rank (L2R), a widely used approach in information retrieval systems aimed at ranking a set of documents based on a user query \cite{hastagger+}.
Gao et al. proposed a hybrid recommendation system that uses a deep neural network combining content-based and collaborative filtering, enhanced with user interest tags and topics to extract heterogeneous features and boost recommendation accuracy and diversity \cite{gao2021hybrid}. Jeong et al. introduced DemoHash \cite{jeong2022demohash}, a multimodal model for personalized hashtag recommendation that incorporates demographic data from user selfies along with textual and visual content.

\smallskip \noindent\textbf{Semi-supervised Models.}
Approaches in this category combine supervised and unsupervised techniques to guide the learning process. For instance, the HASHET model (HAshtag recommendation using Sentence-to-Hashtag Embedding Translation) \cite{HASHET-TKDD-2022} leverages BERT (\textit{Bidirectional Encoder Representations from Transformers}) \cite{devlin2018bert} to compute the embedded representation of a post, and a CBoW (\textit{Continuous Bag of Words}) Word2Vec model \cite{mikolov2013efficient} to determine the latent representation of its hashtags, by capturing semantic and syntactic similarities in an unsupervised manner.
A semantic mapping is then learned via transfer learning as the translation between the BERT-based embedding of a post and the latent representation of its hashtags in the Word2Vec space. This is achieved by stacking a projection head on the BERT encoder and fine-tuning the model in a supervised manner. At inference time, HASHET maps a post into the hashtag embedding space and retrieves the $k$ nearest hashtags to be recommended using cosine similarity.
Unlike other deep learning-based techniques, the recommendation process performed by HASHET relies on the distributional assumption that semantically similar hashtags generate nearby embeddings. This locality concept enables the model to exploit the topic-based clustering structure within the hashtag embedding space, reflecting the learned semantic affinities among hashtags.

\section{Proposed Methodology}
\label{sec:h-adapts}

In this section, we provide a detailed description of the proposed methodology, namely \textit{H-ADAPTS (Hashtag recommendAtion by Detecting and adAPting to Trend Shifts)}, which is specifically designed to recommend relevant and up-to-date hashtags to social media users in real-world dynamic contexts. In particular, H-ADAPTS extends the HASHET model by introducing the ability to cope with trend shifts in social media conversations.
To this end, the Apache Storm framework is leveraged for the continuous monitoring of the unbounded stream of social media posts, enabling the real-time detection of trend shifts in online conversations. In addition, an effective adaptation strategy is introduced to realign the model with the latest trends, thereby keeping pace with newly emerged hashtags, topics, and socially impactful events.

Another key difference from the original HASHET model lies in the use of the \textit{DistilBERT} \cite{sanh2019distilbert} transformer-based encoder, a smaller, faster, and more cost-effective version of BERT derived via teacher-student knowledge distillation \cite{hinton2015distilling}. Compared to the standard BERT encoder used in HASHET, DistilBERT significantly reduces training time by approximately 60\%, making it suitable for ensuring rapid adaptation in dynamic settings. This reduction in training time is crucial, as it allows for quick model realignment following the detection of trend shifts. While this realignment process is performed asynchronously to maintain continuous model availability for user queries, the faster alignment enabled by DistilBERT ensures that the system can more promptly leverage the updated model, thereby enhancing the overall quality of recommendations. Furthermore, DistilBERT offers lower inference time compared to BERT, allowing for faster response times to user queries with minimal loss in accuracy, as extensively demonstrated in the literature \cite{sanh2019distilbert, silalahi2022named, gupta2021leveraging}.

\begin{figure}[ht]
	\centering
	{\includegraphics[width=\linewidth]{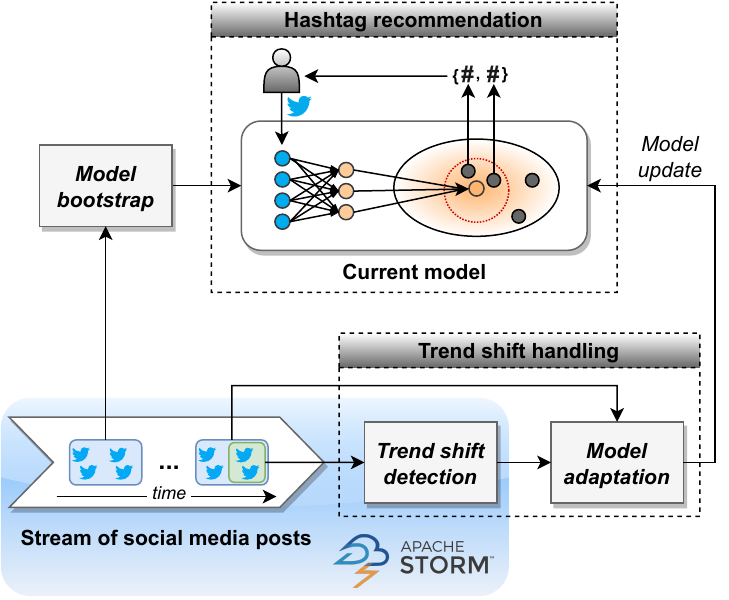}}
	\caption{Execution flow of H-ADAPTS comprising four steps: $(1)$ \textit{model bootstrap}, $(2)$ \textit{trend shift detection}, $(3)$ \textit{model adaptation}, and $(4)$ \textit{hashtag recommendation}.}
	\label{fig:ada-hashet}	
\end{figure}

The main execution flow of H-ADAPTS is summarized in Figure \ref{fig:ada-hashet}, and is divided into four steps:
\begin{enumerate}[leftmargin=*]
    \item \textit{Model bootstrap}: in this phase, all necessary components are initialized, including the inner recommendation model.
    \item \textit{Trend shift detection}: the real-time stream of social posts is processed by Storm to detect a trend shift, i.e., a significant deviation of current online conversation from previous history in terms of main trends and topics.
    \item \textit{Model adaptation}: if a trend shift is detected, the current recommendation model is asynchronously updated, realigning it with the current trends and topics.
    \item \textit{Hashtag recommendation}: in this step, the current recommendation model is used for recommending a set of hashtags for a query post provided by the user.
\end{enumerate}

In what follows, we provide an in-depth description of the different steps that make up the proposed \textit{trend-aware} dynamic hashtag recommendation methodology. For the sake of clarity, Table~\ref{tab:meaning} reports the meaning of the main symbols used throughout the paper. Furthermore, an implementation of H-ADAPTS is publicly available on GitHub\footnote{\url{https://github.com/SCAlabUnical/H-ADAPTS}}.

\begin{table}[h]
	\centering
	\resizebox{0.98\linewidth}{!}{%
		\begin{tabular}{@{}ll@{}}
			\toprule
			\textbf{Symbol} & \textbf{Meaning} \\ \midrule
			$E$ & The pre-trained BERT-based encoder.\\
			$S_{emb}$ & Sentence embedding space.\\
			$W2V$ & The word embedding model, based on Word2Vec.\\
			$W_{emb}$ & Word embedding space.\\
			$MLP$ & The mapper $S_{emb} \rightarrow W_{emb}$, based on a MLP. \\
			$SM$ & The semantic mapping model, i.e., $stack(E, MLP)$.\\
			$\mathcal{M}$ & The HASHET model, defined as $\langle W2V, SM\rangle$.\\
			$\mathcal{S}$ & Unbounded stream of social media posts. \\
			$B$ & The bootstrap window.\\
			$T$ & The current tumbling window, $T \subset W \land T \subset F$.\\
            $W$ & The current sliding window.\\
			$F$ & The current fine-tuning window, $F \subset W$.\\
			$d_{B,T,W,F}$ & Number of days in the $B$, $T$, $W$, and $F$ windows.\\
			$\mathcal{H}^*$ & The current main trends and topics.\\
			$\mathcal{H}^T$ & Top-$n$ hashtags of the posts belonging to $T$.\\
			$\delta$\,=\,$RJD(\mathcal{H}^*, \mathcal{H}^T)$ & Ranked Jaccard Distance between $\mathcal{H}^*$ and $\mathcal{H}^T$.\\
			$\omega$ & The threshold used in the trend shift detection step.\\
			$\mathcal{R}^k_p$ & The set of $k$ hashtags recommended by the model for a post $p$.\\
            $\mathcal{T}^k_p$ & The set of $k$ target hashtags to be recommended for a post $p$.\\
			\bottomrule
		\end{tabular}%
	}
	\caption{Meaning of the main symbols used in the paper.}
	\label{tab:meaning}
\end{table}

\subsection{Trend-Aware Dynamic Hashtag Recommendation}
\label{sec:method_general}

H-ADAPTS effectively deals with the presence of trend shifts by being fully aware of how social trends underlying online conversation vary over time. We treat trend shifts as concept drifts, which means that, from a recommendation perspective, a change in the major social trends driving online conversation can lead to a huge change in the patterns that link a given post to the hashtags that most fit with it.
The proposed methodology performs a windowed adaptation of all its components, that are continuously realigned to the latest trends when a significant shift is detected.
Specifically, a sliding window $W$ is used to maintain the posts generated in the last $d_W$ days (i.e., the recent history), while a daily tumbling window $T$ is used for real-time monitoring of the social data stream to detect a trend shift. In addition, a trending set $\mathcal{H}^*$ is used to maintain a constantly updated representation of the main trends and topics on which social media conversation is currently focusing.

Algorithm \ref{alg:adaptive} shows how the H-ADAPTS model works given an unbounded stream of social media posts. A detailed description of its main steps, devoted to model bootstrap, shift detection, and model adaptation is provided in the following.

\setlength{\algomargin}{1em}
\begin{algorithm}[!htb]
       \fontsize{8pt}{10pt}\selectfont 
	\caption{H-ADAPTS}
	\label{alg:adaptive}
	\DontPrintSemicolon 
	\KwIn{{Windowed stream $\mathcal{S}$, current date $d$, threshold $\omega$, trending set cardinality $n$, size (in days) of the bootstrap, tumbling, sliding, and fine-tuning windows $d_B$, $d_T$, $d_W$, $d_F$}}
	\medskip\tcc{Model bootstrap}
    \smallskip
	$B \gets \mathcal{S}.getLastWindow(d_B, d)$\\
    $W2V \gets Word2Vec.train(B)$\\
	$targets \gets compute\_targets(W2V)$\\
	$E \gets init\_from\_pretrained()$\\
	$MLP \gets init\_from\_scratch()$\\
	$SM \gets stack(E, MLP)$\\
	$SM.transfer\_learning(B, targets)$\hspace{0.1cm} \tcp{train $MLP$ ($E$ is frozen)}
	$SM.fine\_tuning(B, targets)$\hspace{0.1cm} \tcp{unfreeze $E$ to fully fine-tune $SM$}
    $\mathcal{M} \gets \langle W2V, SM \rangle$ \hspace{0.1cm} \tcp{bootstrapped $HASHET$ model}
    $\mathcal{H}^*\gets top\_hashtags(B, n)$\\
	\medskip\tcc{Trend shift handling}
    \smallskip
	\While{$True$}{
	    $d\gets d+d_T$\\
	    \smallskip\tcc{Trend shift detection}
        \smallskip
	    $T \gets \mathcal{S}.getLastWindow(d_T, d)$\\
		$\mathcal{H}^T\gets top\_hashtags(T, n)$\\
		\If{$\delta(\mathcal{H}^*, \mathcal{H}^T) \geq \omega$}{
		\smallskip\tcc{Model adaptation}
        \smallskip
		$W \gets \mathcal{S}.getLastWindow(d_W, d)$\\
		$W2V \gets Word2Vec.train(W)$\\
		$targets \gets compute\_targets(W2V)$\\
		$MLP.reset\_weights()$ \hspace{0.1cm} \tcp{re-initialize the $MLP$ mapper}
		$SM \gets stack(E, MLP)$\\
	    $SM.transfer\_learning(W, targets)$ \hspace{0.1cm} \tcp{re-training of $MLP$}
        $F \gets \mathcal{S}.getLastWindow(d_F, d)$\\
	    $SM.fine\_tuning(F, targets)$ \hspace{0.1cm} \tcp{progressive fine-tuning of $SM$}
	    $\mathcal{M} \gets \langle W2V, SM \rangle$        \hspace{0.1cm} \tcp{updated $HASHET$ model}
	    $\mathcal{H}^*\gets \mathcal{H}^T$        \hspace{0.1cm} \tcp{update current trends}

		}
	}
\end{algorithm}

\smallskip \noindent\textbf{Model Bootstrap.}
In this phase, the HASHET model used within H-ADAPTS is trained on the social media posts belonging to the bootstrap window $B$, which comprises the last $d_B$ days including the current day $d$ (lines 1-10). In addition, the trending set $\mathcal{H}^*$ is initialized with the top-$n$ hashtags of $B$, ordered by decreasing occurrence (line 11). The model is trained through a multi-step process that involves: \textit{(i)} training a Word2Vec model to generate the latent targets for semantic mapping (lines 3-4); \textit{(ii)} training a semantic mapping model, to learn how to map the semantic representation of input posts to the corresponding latent vector in the hashtag embedding space (lines 5-10). The semantic mapping model $SM$ is obtained by stacking a multilayer perceptron, i.e., the $MLP$ mapper, on top of the pre-trained BERT encoder $E$ (lines 5-7). It is trained in two steps:
\begin{itemize}[leftmargin=*]
    \item \textit{Transfer learning} (line 8). In this step, the BERT encoder $E$ is frozen and used as a feature extractor, to compute a latent representation of the input posts as the global average pooling over the embedded representation of words. The $MLP$ mapper then translates each representation into a target vector lying in the latent space of hashtags. It is built as a multi-layer perceptron trained from scratch with a cosine distance loss, which measures, for a given input post, the distance between the predicted vector and the true target, defined as the average embedding of the hashtags contained in that post.
    \item \textit{Fine tuning} (line 9). The entire semantic mapping model $SM$, composed of the unfrozen BERT encoder $E$ and the mapper $MLP$, is fully fine-tuned to incrementally adapt the pre-trained features of the encoder to the translation task, thus refining BERT-generated embeddings to facilitate their translation into the hashtag embedding space. Furthermore, in this step, a low learning rate is used to prevent pre-trained features from being distorted by large weight updates.
\end{itemize}

\smallskip \noindent\textbf{Trend Shift Detection.}
Once bootstrapped, the model is ready to suggest hashtags to users. However, the quality of its recommendations is likely to deteriorate over time due to significant changes in online conversations, driven by the emergence of new hashtags that reflect trending topics and socially impactful events. As mentioned above, we treat these shifts as concept drifts, since from a recommendation perspective, they can cause substantial misalignment in the current model by altering the patterns that link social posts to hashtags.

To address this issue, H-ADAPTS performs a real-time trend shift detection step, which relies on the identification of a significant deviation in the current trends of online conversation. Then, if a shift is detected, the model can dynamically realign with the latest trends.
In particular, given the current tumbling window $T$, consisting of the last $d_T$ days including the current date $d$, the set $\mathcal{H}^T$ containing the top-$n$ hashtags of $T$ is compared with the $\mathcal{H}^*$ set, which stores the top-$n$ relevant hashtags and acts as a representation of the current trends underlying social media conversation (lines 14-17). To measure the dissimilarity between these two rankings, we introduce the \textit{Ranked Jaccard Distance (RJD)} metric, a variation of the Jaccard Distance we designed to measure the distance between ranked sets. As the classical Jaccard index, this metric is defined from the concepts of interception and union. Let $rank(S,h) = n-i$ be the rank of hashtag $h$ in a ranked set $S$, where $n = |S|$ is the maximum assignable rank and $i$ is the position of $h$ in the ranking. Consequently, hashtags in the first positions are given a higher rank. We define the ranked intersection between two rankings $\mathcal{H}^\prime$ and $\mathcal{H}^{\prime\prime}$ as follows:
\vspace{-0.05cm}
\begin{equation*}
rank(\mathcal{H}^\prime \cap \mathcal{H}^{\prime\prime}) = \sum\limits_{h \in \mathcal{H}^\prime \cap \mathcal{H}^{\prime\prime}}\min{\{rank(\mathcal{H}^\prime,h), rank(\mathcal{H}^{\prime\prime}, h)\}}
\end{equation*}

In this formula, instead of counting the number of hashtags in the intersection, we sum up a score for each hashtag $h$, computed as the minimum rank of $h$ in the two sets.
Differently, starting from the set representing the union of the hashtags in the two ranks, the ranked union is computed by summing up a score for each hashtag $h$ in the union set, defined as the average rank of $h$ in the two sets. Formally:
\vspace{-0.01cm}
\begin{equation*}
rank(\mathcal{H}^\prime\cup \mathcal{H}^{\prime\prime}) = \sum\limits_{h \in \mathcal{H}^\prime \cup \mathcal{H}^{\prime\prime}}\frac{rank(\mathcal{H}^\prime,h) + rank(\mathcal{H}^{\prime\prime}, h)}{2}
\end{equation*}

Finally, similarly to the standard Jaccard distance, the proposed ranked variation is defined as:
\vspace{-0.01cm}
\begin{equation}
RJD(\mathcal{H}^\prime, \mathcal{H}^{\prime\prime})=1 - \frac{rank(\mathcal{H}^\prime \cap \mathcal{H}^{\prime\prime})}{rank(\mathcal{H}^{\prime}\cup \mathcal{H}^{\prime\prime})}
\end{equation}

If the measured distance (i.e., $\delta$) exceeds a predetermined threshold $\omega$ (line 18), a trend shift is detected, triggering the adaptation step. This allows the model to realign with emerging trends, which may include newly appeared hashtags or previously encountered ones that have regained relevance due to specific events capturing public attention. In either case, the learned relationships linking the semantics of the posts to the associated hashtags must be updated to ensure high-quality, up-to-date recommendations.

\smallskip \noindent\textbf{Model Adaptation.}
The adaptation process (lines 19-29) is necessary to extend the knowledge of the inner recommendation model used within H-ADAPTS to new emerging trends, incorporating unknown hashtags and understanding how already known ones are used in different contexts, based on the latest discussion topics. In particular, when a trend shift is detected, the different parts that make up the recommendation model (i.e., the latest updated HASHET model) are asynchronously updated in three steps:
    \begin{itemize}[leftmargin=*]
        \item \textit{Update of the hashtag embedding space}.
        This step (lines 20-21) allows the model to discover and understand the contextual relationships between words and trending hashtags currently used by social users. Specifically, the Word2Vec model is trained on the tweets within the current sliding window $W$, allowing the system to map previously unknown hashtags to specific concepts and capture semantic shifts in already known hashtags. This ensures that the underlying topic-based clustering structure remains aligned with the current usage of hashtags.
        \item \textit{Update of the MLP mapper}. In this step, the projection head of the semantic mapping model, i.e., the $MLP$ mapper stacked on top of the BERT encoder $E$, is adapted to the new hashtag embedding space (lines 22-25). In particular, the mapper is re-initialized from scratch (line 23) and is trained via transfer learning on the social posts within the current $W$ window (lines 24-25). The target vectors are derived from the updated hashtag embedding space generated in the previous step (line 22), while the semantic embeddings are produced using the (frozen) BERT encoder from the current semantic mapping model. This training phase, driven by a cosine distance loss, allows for a smooth transition to newly discovered concepts by maximizing semantic similarity between the mapper’s predicted vectors and the actual target vectors in the updated hashtag space.
        \item \textit{Progressive fine-tuning of the semantic mapping model}.
        In this step, the BERT encoder is unfrozen and the whole semantic mapping model $SM$ is fully fine-tuned with a low learning rate, starting from its current weights (lines 26-27). This step is performed using the post within a fine-tuning window $F \subset W$, which includes the most recent $d_F$ days, with $d_F << d_W$. Through this step, the BERT encoder can be smoothly adapted progressively, thus generating more suitable embeddings to be fed to the mapper for translation into the newly generated hashtag embedding space. Additionally, the mapper is concurrently adapted to these fine-tuned semantic representations, enhancing the overall cohesion of the semantic mapping model and improving the accuracy of hashtag recommendations.
    \end{itemize}

Lastly, the updated model $\mathcal{M}$ is obtained by joining the adapted $W2V$ and $SM$ models, and the $\mathcal{H}^*$ set is updated with the top-$n$ hashtags of the current tumbling window $T$, i.e., the set $\mathcal{H}^T$ (lines 28-29).

\smallskip \noindent\textbf{Hashtag Recommendation.}
In this step, the latest updated HASHET model $\mathcal{M}$, is used for recommending a set of $k+\eta$ representative hashtags for a query post $p$ provided by the user. In particular, given an input post $p$, the semantic mapping model $SM$ is leveraged to compute the corresponding target vector $v^*_p$ in the hashtag embedding space. Next, the set of hashtags to recommend $\mathcal{R}^k_p$ is found, by identifying the $k$ nearest hashtags of $v^*_p$ in the hashtag embedding space, ordered by decreasing cosine similarity.
Finally, the $k$ nearest hashtags search is extended by $\eta$ steps, to include additional hashtags that share semantic context with the target vector, thus capturing semantic equivalences. This hashtag recommendation process benefits from the adaptive nature of H-ADAPTS, enabling the model to align with evolving social trends and ultimately improving recommendation accuracy over time.

\section{Storm-based System Design}
\label{sec:deployment}

\begin{figure}[ht]
	\centering
	{\includegraphics[width=\linewidth]{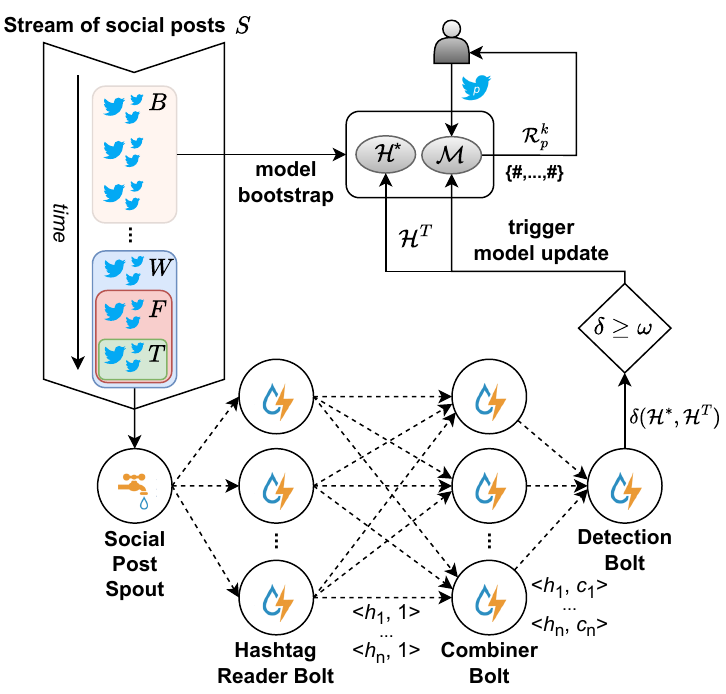}}
	\caption{Storm topology supporting the execution of H-ADAPTS given an unbounded stream of social media posts.}
	\label{fig:storm-topology}	
\end{figure}

A key feature of H-ADAPTS is the real-time detection of trend shifts in online conversation, which is achieved by using Storm, a real-time computation system that allows the fast and reliable processing of unbounded data streams. Figure \ref{fig:storm-topology} illustrates the whole Storm topology, highlighting how it interacts with the unbounded stream of tweets to enable real-time detection of trend shifts and model adaptation.

\subsection{Why Choosing Storm} 
As mentioned before, our methodology leverages Apache Storm to efficiently manage unbounded streams of social media data, which is key for the effective real-time detection of the main trend shifts in online conversation. This framework has been widely used in the literature to enable the low latency, fault-tolerant, and scalable analysis of social media data streams \cite{amen2022big, sharma2021recent, gupta2018unleashing, Elbedwehy2022}, which are crucial properties for real-world dynamic scenarios. Indeed, compared to other frameworks like Apache Spark Streaming, Storm effectively handles high-volume and high-velocity data streams while maintaining high availability. Conversely, Spark Streaming relies on micro-batch processing, resulting in lower performance compared to pure streaming processing systems, like Storm and Apache Flink, as discussed in various benchmarks \cite{7530084, 8509390, 7841533}. Flink and Storm exhibit similar performances, but Storm provides a more mature and robust ecosystem encompassing tools and libraries for data ingestion, processing, storage, monitoring, and troubleshooting \cite{belcastro2022programming}, which simplify the development, testing, and deployment of the real-time hashtag recommendation system. Furthermore, Storm's multi-language protocol enables flexibility in the choice of programming language, allowing the incorporation of Python scripts within a Java-based topology. This facilitates the integration of multiple independent modules, enhancing modularity and code reuse while harnessing Storm's real-time analysis capabilities. Hence, this feature is key for incorporating the HASHET model into a broader framework, enhancing it with trend shift awareness and adaptiveness.

\subsection{Topology Design}
The real-time logic of the system is enclosed within a Storm topology consisting of three main components:

\begin{itemize}[leftmargin=*]
    \item A \textit{Social Post Spout}, which collects social media posts and emits them into the topology, potentially filtering them based on certain criteria (e.g., keywords, location, and language). These social media posts are then emitted to the next bolts as tuples for further processing.
    \item A \textit{Hashtag Reader Bolt}, which receives social media posts from the Social Post Spout and extracts hashtags by using regular expressions and text processing techniques. Specifically, for each hashtag $h$, it emits a tuple $(h, 1)$ to the next bolt, to mark the occurrence of that hashtag.
    \item A \textit{Detection Bolt}, which detects trend shifts and triggers the adaptation of the HASHET model to current trends. Particularly, it counts hashtag occurrences from the Hashtag Reader Bolt within a tumbling window $T$ and identifies trending hashtags, which are compared with current trends to detect significant shifts. Upon detecting a trend shift, the bolt triggers model adaptation using data from the last sliding windows $W$ and $F$, with $F \subset W$.
\end{itemize}
Therefore, the logic of the trend shift detection is enclosed in the Detection Bolt, which compares the $\mathcal{H}^T$ set containing the top-$n$ hashtags in the current tumbling window $T$ with the current trending set $\mathcal{H}^*$. In particular, given a hashtag $h$, the bolt receives $m$ pairs $(h$, $1)$ from the Hashtag Reader Bolt, where $m$ is the number of occurrences of $h$ in $T$. Then, these pairs are aggregated, generating as output a pair $(h$, $m)$ for each hashtag $h$, thus obtaining the $\mathcal{H}^T$ set. By measuring the Ranked Jaccard Distance between $\mathcal{H}^*$ and $\mathcal{H}^T$, H-ADAPTS automatically determines if the main trends have changed to such an extent that a realignment of the current recommendation model is required, i.e., a trend shift has occurred. This decision, as explained earlier, is controlled by a hyperparameter $\omega$, which is a threshold for the distance $RJD(\mathcal{H}^*, \mathcal{H}^T)$, specifying the maximum deviation from the current trends beyond which a realignment is necessary.

\smallskip \noindent\textbf{Use of Combiners and Stream Grouping.}
To ensure the high efficiency of the Storm topology, reducing the workload of the Detection bolt is crucial. One effective approach is to introduce a \textit{Combiner} or mini-reducer bolt before the Detection Bolt. This Combiner aggregates the information from Hashtag Reader bolts with the same key (i.e., the same hashtag) before passing it to the Detection Bolt, significantly reducing the number of tuples processed by the Detection Bolt and thus alleviating the bottleneck in the entire topology. Efficient grouping strategies are also key for real-time system performance since they determine how the stream is distributed among different tasks. \textit{Shuffle grouping} randomly distributes tuples across worker processes, preventing any single worker from being overloaded. \textit{Field grouping} routes tuples based on one or more fields, directing tuples with the same field value to the same worker process. In the real-time social media processing scenario addressed in this work, hashtags exhibit a highly skewed distribution, with a small number of extremely popular hashtags and a large number of infrequently used ones. Therefore, field grouping may overload some workers while leaving others underutilized. Conversely, shuffle grouping achieves a balanced workload distribution among available worker processes, enhancing the overall performance of the whole Storm topology.

\section{Experimental Evaluation}
\label{sec:performance}

In this section, we describe the extensive experimental evaluation we carried out to assess the effectiveness of H-ADAPTS in recommending relevant hashtags to social media posts. Particular attention is paid to the ability of our methodology to detect trend shifts in real time, adapting to them through a realignment process. We compared our realignment strategy, which involves retraining the projection head (MLP mapper) and continuous fine-tuning of the entire semantic mapping model, with other possible strategies to highlight the main advantages of the selected approach.
All the experiments were performed on two real-world case studies, related to the COVID-19 global pandemic and the 2020 US presidential election, respectively. For each case study, we demonstrate how the model effectively detects and adapts to trend shifts, maintaining superior performance compared to competing techniques, even in the presence of emerging topics and newly introduced hashtags. Furthermore, all identified shifts are analyzed to provide insights into the new trending topics and hashtags discovered by the model.

\subsection{Experimental Settings}
\label{ssec:exp_set}

This section outlines the experimental settings used in the two case studies, including the evaluation metrics, the hyperparameters applied, and the baseline techniques selected for comparison. All experiments were conducted on a high-performance computing system running the Linux operating system, equipped with an Intel Xeon Gold 6248R CPU, eight NVIDIA A30 GPUs, and 754 GB of RAM.

\smallskip \noindent\textbf{Evaluation metrics.} 
A rank-based version of the recall measure was used to evaluate the performance of the proposed model. Given a post $p$ and the set of its target hashtags $\mathcal{T}^k_p$, the model outputs the set of recommended hashtags $\mathcal{R}^k_p=\{r^1_p, r^2_p,\dots, r^k_p\}$. To determine the relevance of each recommended hashtag for the post $p$, we defined a function $rel(r^i_p, p)$ as follows:
\begin{equation}
rel(r^i_p, p)=\begin{cases}
1 \hspace{0.2cm} \text{if }r^i_p\in \mathcal{T}^k_p\\
0 \hspace{0.2cm} \text{otherwise}\\
\end{cases}\hspace{-0.35cm},\,\forall i\in \{1, \dots, k\}
\end{equation}

In other words, $rel(r^i_p, p)=1$ if the recommended hashtag is relevant for $p$, i.e., it is a target hashtag that should be recommended by the model.
Using this relevance function, the recall measure $R@k$ can be expressed as follows:
\begin{equation}
R@k(p)=\frac{1}{\lvert \mathcal{T}^k_p \lvert}\sum_{i=1}^{k}rel(r^i_p, p)
\end{equation}
It represents the model's hit rate and is calculated as the fraction of target hashtags that were successfully recommended.

\smallskip \noindent\textbf{Hyperparameter Setting.} We set the value of the main hyperparameters used during the experimental evaluation as follows. The length of the bootstrap window ($B$), denoted by $d_B$, is fixed at two weeks for all techniques. The length of the tumbling window ($T$), denoted by $d_T$, is set to one day, meaning that the methodology analyzes the hashtags of the current day to detect a trend shift. The length of the sliding window ($W$), denoted by $d_W$, is set to two weeks to maintain recent history. The length of the fine-tuning window ($F$), denoted by $d_F$, is set to four days. The RJD threshold $\omega$ used for trend shift detection is set to 0.9. The size of the trending set of hashtags $\mathcal{H}$, denoted by $n$, is set to 10. Finally, to ensure a fair comparison with competing techniques, which do not perform semantic expansion, the $\eta$ factor is set to 0.

\smallskip \noindent\textbf{Selected Techniques for Comparison.}
To assess the effectiveness of H-ADAPTS in recommending relevant hashtags and adapting to trend shifts over time, we conducted a thorough comparison with various techniques in the literature.
In particular, we compared with unsupervised techniques, following generative (i.e., \textit{LDA-GIBBS} \cite{godin2013topic}), frequency-based (i.e., \textit{HF-IHU} \cite{otsuka2016hashtag}), and clustering-based (i.e., \textit{W2V+DBSCAN} \cite{ben-lhachemi2018}) approaches.
Moreover, we compared with supervised deep neural models that rely on the attention mechanism, i.e., the Topical Co-Attention Network \textit{TCAN} \cite{li2019topical}, and its degenerate version \textit{GGA-BLSTM}, in which the model can only attend to the semantic content of the post, without any topical information.
We also evaluated the performance of H-ADAPTS against a fully fine-tuned BERT classifier, obtained by replacing the projection head of HASHET with a multi-class classification head. In this model, a softmax activation is used to distribute the probability over all possible candidate hashtags, following the approach proposed in \cite{mahajan2018exploring}, instead of exploiting locality in the hashtag embedding space. Lastly, an ablation analysis is conducted through a point-wise comparison between H-ADAPTS and the standalone HASHET model, to gain insights into the advantages of introducing trend shift awareness.

It is important to note that all techniques selected for comparison were implemented without any modifications specifically tailored for real-time scenarios, such as periodic retraining, as seen in other studies \cite{hastagger+}. Incorporating such mechanisms would introduce additional hyperparameters to control the retraining process, which may be difficult to tune and highly dependent on the specific application context. Moreover, relying on a fixed retraining frequency can lead to significant inefficiencies, either by triggering unnecessary updates or by missing critical ones.

\subsection{COVID-19 Pandemic}

This section presents the analysis carried out using H-ADAPTS on a corpus of $685,284$ social posts from $239,926$ users related the COVID-19 pandemic, published on X (formerly Twitter) between August 1, 2020 and September 30, 2020. The posts were filtered based on specific keywords related to the COVID-19 pandemic, such as ``COVID", ``coronavirus", ``pandemic", and ``lockdown". In particular, as detailed in Section \ref{ssec:exp_set}, we used the first two weeks---from August 1 to August 14---to bootstrap H-ADAPTS and train all other models selected for comparison. Then, the subsequent weeks until September 30 were used to test the quality of the recommendations generated by the models over time, in a real-time fashion. It is worth noticing that, up until the first adaptation occurs, H-ADAPTS and the standalone HASHET model produce identical outputs, while, after the first adaptation, their behaviors diverge consistently.

\smallskip \noindent\textbf{Comparison of Model Adaptation Strategies.}
As detailed in Section \ref{sec:method_general}, the adaptation process involves updating the words/hashtags embedding model $W2V$, learned by Word2Vec, and the semantic mapping model, composed of the DistilBERT encoder $E$ and the $MLP$ mapper stacked on top of $E$. As regards the update of the hashtag embedding space, it is essential to realign the Word2Vec model $W2V$ using the tweets belonging to the current sliding window $W$, to handle unknown hashtags and intercept the changes in the semantics of already known hashtags. In this way, the clustering structure underlying the hashtag embedding space is realigned to the current usage of hashtags in the online conversation. On the contrary, considering the update of the semantic mapping model, there may be alternative strategies to perform a realignment, which slightly differ from the approach leveraged by H-ADAPTS. Therefore, we compare the proposed model adaptation strategy with three different possible alternatives. 

Let $TL(\cdot, \cdot)$ and $FT(\cdot, \cdot)$ be two functions representing the transfer learning and fine-tuning operations respectively, and let their arguments be $(i)$ the component involved in the learning process and $(ii)$ the window from which the used data are gathered. According to this notation, the strategies we devised for comparison can be described as follows:
\begin{itemize}
\item TL($MLP$, $W$) + FT($E+MLP$, $W$): both transfer learning and end-to-end fine-tuning are performed on the sliding window $W$. As in the proposed strategy, transfer learning involves the $MLP$ mapper, which is reinitialized from scratch, while fine-tuning is performed on the entire semantic mapping model $E+MLP$, with the original weights of the pre-trained encoder $E$ being restored.
\item FT($MLP$, $F$): it only performs the progressive fine-tuning of the $MLP$ mapper on the fine-tuning window $F$, with the encoder serving as a feature extractor.
\item FT($E+MLP$, $F$): the entire semantic mapping model $E+MLP$ is fine-tuned progressively in an end-to-end manner on the tweets within the fine-tuning window $F$, without prior transfer learning on the $MLP$ mapper. \end{itemize}

\begin{figure}[ht]
	\centering
	{\includegraphics[width=\linewidth]{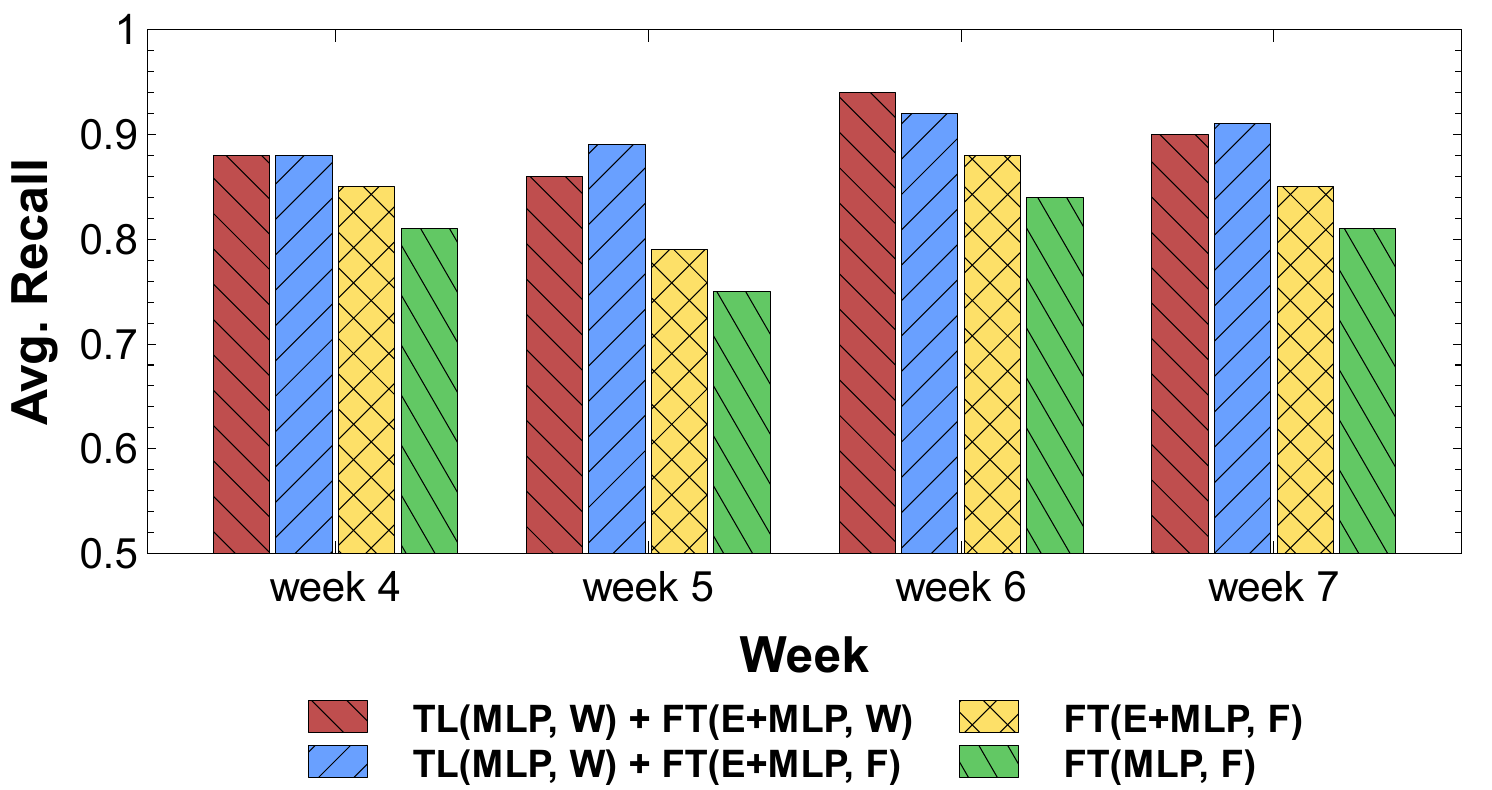}}
	\caption{Comparison of weekly average recall among the different alternative strategies and the proposed one, i.e., TL($MLP$, $W$) + FT($E$$+$$MLP$, $F$).}
	\label{fig:retrain_strat}	
\end{figure}

Figure \ref{fig:retrain_strat} reports a comparison of weekly average recall among the different alternative strategies and the proposed one, which can be formalized as TL($MLP$, $W$) + FT($E$$+$$MLP$, $F$), according to the notation introduced above. The comparison is provided starting from the fourth week, during which the first shift is detected, to observe how recommendation performance varies with the use of different adaptation strategies. In addition, Table \ref{tab:comp_eff} compares these strategies in terms of recommendation performance, averaged across all weeks, and computational efficiency, measured by the duration of a single training epoch, memory usage, and energy consumption.

\begin{table}[h]
\centering
\resizebox{\linewidth}{!}{
\begin{tabular}{@{}c|c|cccc@{}}
\toprule
\begin{tabular}[c]{@{}c@{}}\textbf{Model Adaptation}\\ \textbf{Strategy}\end{tabular} & \begin{tabular}[c]{@{}c@{}}\textbf{Avg.}\\ \textbf{Recall}\end{tabular}
&\begin{tabular}[c]{@{}c@{}}\textbf{Duration}\\ \textbf{(sec.)}\end{tabular} & 
\begin{tabular}[c]{@{}c@{}}\textbf{Memory}\\ \textbf{(MB)}\end{tabular} &
\begin{tabular}[c]{@{}c@{}}\textbf{Energy}\\ \textbf{(KWh)}\end{tabular} &
\begin{tabular}[c]{@{}c@{}}\end{tabular} \\
\midrule
TL(MLP, W) + FT(E+MLP, W) & 0.90 & 730.68 & 899.61 & $13.95 \cdot 10^{-2}$\\
\textbf{TL(MLP, W) + FT(E+MLP, F)} & 0.90 & 387.97 & 880.40 & $7.26 \cdot 10^{-2}$ \\
FT(E+MLP, F)              & 0.84 & 145.13 & 190.85 & $2.69 \cdot 10^{-2}$ \\
FT(MLP, F)                & 0.80 & 71.51  & 190.91 & $1.30 \cdot 10^{-2}$ \\
\bottomrule
\end{tabular}%
}
\caption{Comparison of recommendation performance and computational efficiency among the different alternative strategies and the proposed one (in bold).}
\label{tab:comp_eff}
\end{table}

As can be seen from Figure \ref{fig:retrain_strat}, the TL($MLP$, $W$) + FT($E$$+$$MLP$, $W$) strategy and the proposed one, i.e. TL($MLP$, $W$) + FT($E$$+$$MLP$, $F$), clearly outperform the other two. However, despite showing comparable performance in terms of average recall, the first strategy uses the whole sliding window $W$, while the proposed one performs a progressive end-to-end fine-tuning on the $F$ window, which is much smaller than $W$ (e.g., in our experiments, $d_F \approx \frac{d_W}{4}$). This progressive fine-tuning allows for a smoother transition of the whole model when adapting to the newly discovered concepts, ensuring high accuracy while nearly halving training time and energy consumption.

The other two strategies, i.e., FT($MLP$, $F$) and FT($E$$+$$MLP$, $F$), only perform the fine-tuning step, without prior transfer learning on the $MLP$ mapper. The difference between them lies in the components whose weights are fine-tuned, i.e., the $MLP$ mapper and the entire stack $E$$+$$MLP$, respectively. The FT($MLP$, $F$) strategy achieves the worst recommendation accuracy due to the lack of strong alignment among the encoder, the mapping head, and the updated hashtag embedding space. Indeed, during the adaptation, the latent space in which the target hashtags are embedded is realigned to incorporate newly emerged hashtags and semantic variations, which requires the realignment of the semantic mapping model. However, the FT($MLP$, $F$) only fine-tunes the current mapping head to the new hashtag embedding space, without fine-tuning the encoder, which is never updated and may remain partially anchored to what was seen during the bootstrap phase. Therefore, although this strategy is the most computationally efficient alternative, as shown in Table \ref{tab:comp_eff}, it causes the encoder to become misaligned with the evolving hashtag embedding space. As a result, the quality of sentence embeddings degrades as new concepts are added, ultimately reducing recommendation effectiveness.

The last strategy, i.e., FT($E$$+$$MLP$, $F$), tries to overcome this issue by fine-tuning the whole semantic mapping model to realign the entire stack, including the encoder. This results in higher recommendation accuracy compared to FT($MLP$, $F$), with only a slight decrease in computational efficiency. However, it remains significantly less accurate than the proposed strategy, which also incorporates a transfer learning phase. By only performing a direct fine-tuning of the whole semantic mapping model, the FT($E$$+$$MLP$, $F$) strategy indeed struggles to adapt effectively to the newly generated hashtag embedding space, as the semantic relationships among latent hashtag representations can vary significantly. In contrast, by incorporating a transfer learning step before fine-tuning, the proposed strategy can lead to better adaptation. Specifically, by initially aligning the $MLP$ mapper to the updated hashtag latent space, this strategy sets a good foundation for later fine-tuning, enabling the subsequent generation of more suitable embeddings to be fed to the updated $MLP$ mapper. In addition, during fine-tuning, the mapper is jointly adapted to these refined embedded representations, leading to the generation of more precise mappings. This results in greater cohesion within the semantic mapping model and improves the model recommendations, making the proposed strategy the optimal choice among all the alternatives discussed, achieving the best trade-off between accuracy and computational efficiency.

\smallskip \noindent\textbf{Detected Trend Shifts and Advantages of Adaptation.}
Here we discuss the main trend shifts identified by H-ADAPTS and demonstrate, through an ablation study, how adapting to these shifts leads to improved recommendation performance. By analyzing the hashtags and topics detected by H-ADAPTS, we identified a macro topic encompassing all COVID-19-related content, which can be further broken down into several micro-topics related to public health guidelines and government policies, such as the effectiveness of mask-wearing, social distancing, and lockdown measures. Our methodology detected two significant trend shifts, each associated with events or phenomena that catalyzed the attention of the online conversation, whose occurrence triggered model adaptation. These shifts, along with the initial knowledge of the model from the bootstrap phase, are described in Table \ref{tab:topics_covid}, which also reports the corresponding topics and the top related hashtags.

\begin{table}[h]
\centering
\resizebox{\linewidth}{!}{
\setlength{\tabcolsep}{3pt}
\begin{tabular}{@{}lll@{}}
\toprule
\multicolumn{1}{c}{\textbf{Start date}} & \multicolumn{1}{c}{\textbf{Topic}} & \multicolumn{1}{c}{\textbf{Top hashtags (trending set)}} \\ \midrule
\vspace{0.15cm}
\begin{tabular}[c]{@{}l@{}}August 1, 2020\\ (\textit{bootstrap phase})\end{tabular} & \begin{tabular}[c]{@{}l@{}}Discussion about\\ COVID-19 and public\\anti-contagion rules\end{tabular} & \begin{tabular}[c]{@{}l@{}}\#covid19, \#coronavirus, \#pandemic, \#wearamask, \\ \#bloodmatters, \#stayhome, \#staysafe, \#sarscov2, \\ \#reallifeheroes, \#washyourhands 
\medskip
\end{tabular} \\
\vspace{0.15cm}
\begin{tabular}[c]{@{}l@{}}Sept. 10, 2020\\(\textit{first shift})\end{tabular} & \begin{tabular}[c]{@{}l@{}} Trump's management\\of COVID-19\\health emergency \end{tabular} & \begin{tabular}[c]{@{}l@{}}\#trumpknew, \#trumphidthetruth, \\ \#deathofdemocracy,  \#covid19, \#trump, \\ \#trumpdoesntcare, \#trumpliedpeopledied, \\ \#trumpvirus, \#trumpliedamericansdied, \#heknew
\medskip
\end{tabular} \\

\begin{tabular}[c]{@{}l@{}}Sept. 24, 2020\\(\textit{second shift})\end{tabular} & \begin{tabular}[c]{@{}l@{}}UNGA event on\\ COVID-19 impact \\ and BTS message\end{tabular} & \begin{tabular}[c]{@{}l@{}}\#covid19, \#coronavirus, \#staysafe, \#wearamask, \\\#pandemic, \#covid, \#unga, \#bts, \#btsonunga, \\\#btsxunga \end{tabular} \\ \bottomrule
\end{tabular}%
\setlength{\tabcolsep}{-3pt}
}
\caption{Main trend shifts detected by H-ADAPTS in the COVID-19 case study.}
\label{tab:topics_covid}
\end{table}

During the first period encompassing the days included in the model bootstrap and test days prior to the first trend shift---from August 15, 2020 to September 9, 2020---there was widespread interest in public health measures to combat COVID-19. Social media users employed hashtags aimed at raising awareness about the pandemic and encouraging people to protect themselves and others from the virus. General hashtags such as \#covid19, \#coronavirus, and \#pandemic were used to discuss the virus and its impact on society, health, and economy. Others, like \#stayhome, \#wearamask, and \#washyourhands were used to promote the adoption of preventive measures like social distancing, mask-wearing, and handwashing to slow the spread of the virus. In addition, the hashtag \#bloodmatters was related to the shortage of blood donations induced by the fear of exposure to the virus, while \#reallifeheroes highlighted the essential role of medical personnel during the crisis.

The first trend shift, detected on September 10, 2020, was related to the actions taken by US President Donald Trump during the pandemic. Specifically, social media users criticized him for being aware of the severity of the virus but lying to the public about its seriousness without taking adequate measures to prevent its spread, resulting in a significant number of deaths in the United States. Detected hashtags associated with these accusations are \#trumpknew, \#heknew, \#trumphidthetruth, \#trumpliedpeopledied, and \#trumpliedamericandied. This topic gained traction following the release of a recorded conversation by The Washington Post on September 9, in which Trump admitted to intentionally minimizing the threat posed by the virus. Additional hashtags like \#deathofdemocracy suggest a broader criticisms of Trump's policies and their perceived impact on American democracy.

The second shift, detected on September 24, 2020, was linked to an event held during the 75th session of the United Nations General Assembly (UNGA), which opened on September 15, 2020. The event focused on the impact of the COVID-19 crisis on future generations, discussing strategies to mitigate its protraction and prepare for a potential second wave. The event gained wide media attention, also due to a video message delivered by the international boy band BTS, with related hashtags including \#btsonunga and \#btsxunga.

\begin{figure}[h]
	\centering
	{\includegraphics[width=\linewidth]{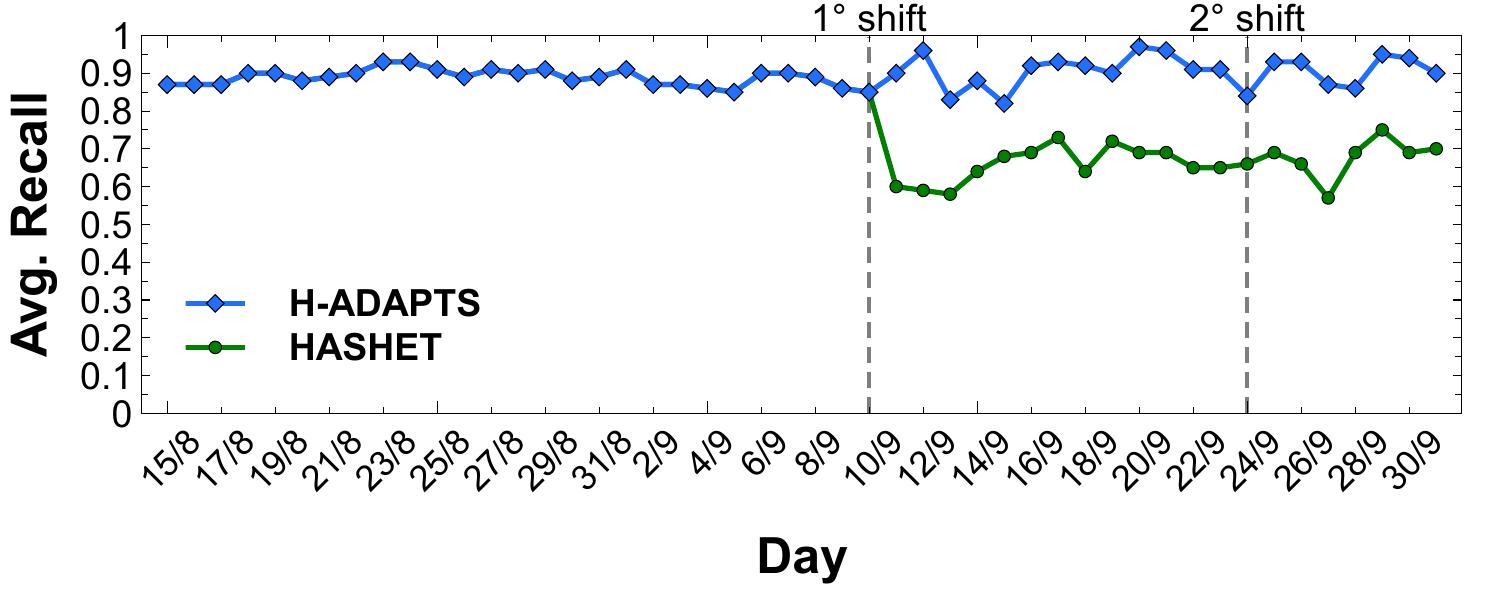}}
	\caption{Point-wise daily comparison between H-ADAPTS and HASHET for the COVID-19 pandemic case study. Trend shifts are indicated by vertical dotted lines.}
	\label{fig:covid_days}	
\end{figure}

This broad range of micro topics reflects the multifaceted and dynamic nature of the social conversation around the COVID-19 pandemic, which must be effectively handled to ensure high-quality recommendations. To better assess this aspect, Figure \ref{fig:covid_days} provides an ablation analysis based on the point-wise comparison of the daily recall rates achieved by H-ADAPTS with and without trend shift awareness. In particular, removing trend shift awareness from H-ADAPTS is equivalent to using a standalone HASHET model, since the inner recommendation model, trained in the bootstrap phase, is never updated over time.

\begin{figure}[ht]
	\centering
	{\includegraphics[width=\linewidth]{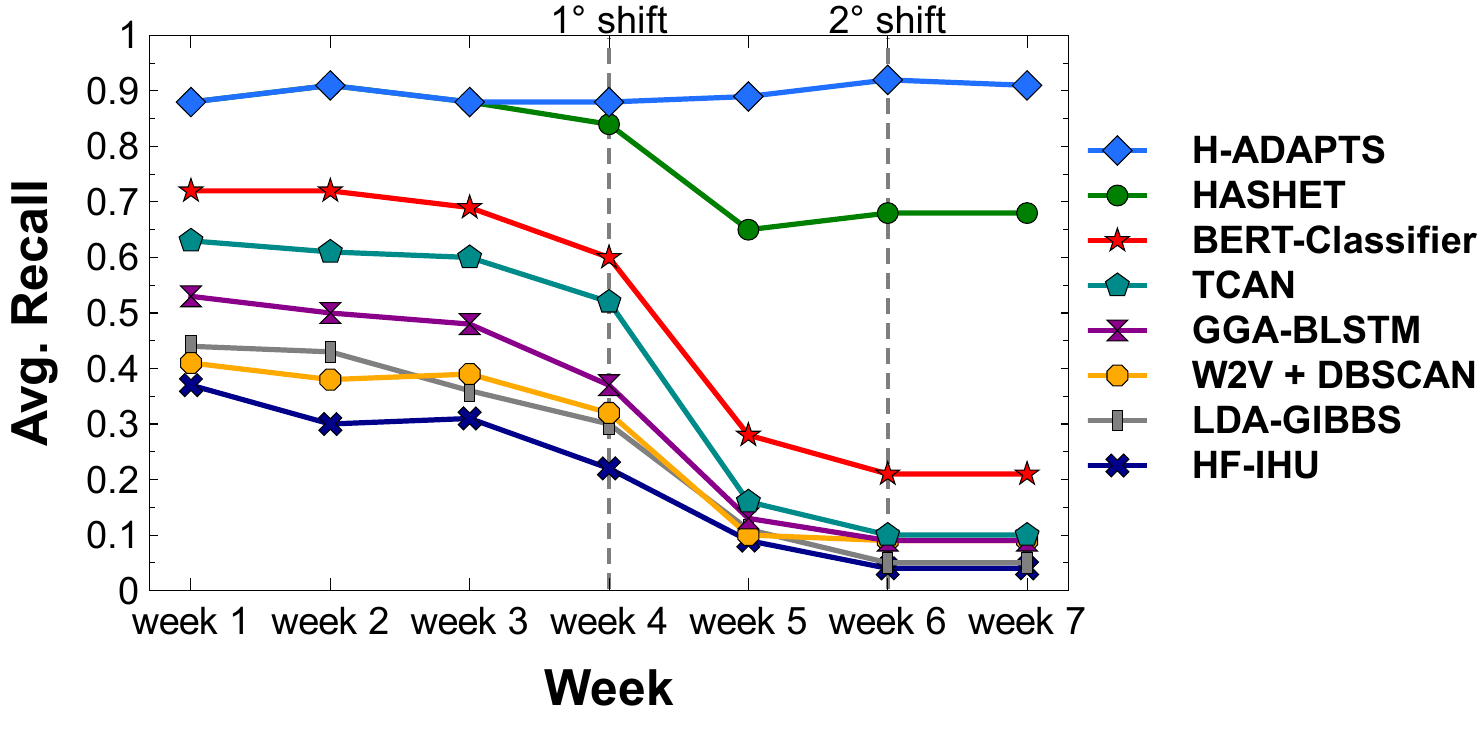}}
	\caption{Comparison with related techniques over time for the COVID-19 pandemic case study, in terms of average recall. Vertical dotted lines indicate trend shifts and corresponding adaptations by H-ADAPTS.
}
	\label{fig:covid_weeks}	
\end{figure}

\smallskip \noindent\textbf{State-of-the-art Comparison.}
In this section, we compare state-of-the-art techniques with H-ADAPTS. Achieved results are depicted in Figure \ref{fig:covid_weeks}, where the trend shifts, detected in the fourth and sixth week, are indicated by vertical dotted lines. As regards test days preceding the first detected shift, the proposed model and the standard HASHET are identical, as no update has been performed yet. 

The recommendation results achieved by the compared techniques can be summarized as follows:
\begin{itemize}[leftmargin=*]
    \item Among unsupervised models, generative (LDA-GIBBS) and clustering-based ones (W2V\,+\,DBSCAN) were able to capture more useful semantic information than simple frequency-based scoring techniques (HF-IHU), leading to more representative suggested hashtags.
    \item Attention-based supervised models (GGA-BLSTM and TCAN) outperformed traditional techniques (HF-IHU, DBSCAN, and LDA) due to their ability to learn a semantically rich representation of the post. However, the topical co-attention model performed slightly better than GGA-BLSTM by jointly using content and topic attention.
    \item The fine-tuned BERT classifier achieved even more accurate results, which is consistent with the effectiveness of transfer learning with large language models.
    \item Lastly, H-ADAPTS and the standalone HASHET, which use a transformer encoder and exploit locality in the hashtag embedding space, outperformed all other techniques.
\end{itemize}

Following the occurrence of trend shifts, with the first one detected during the fourth week, the performance of non-dynamic techniques degrades significantly. Their inability to adapt to newly emerged concepts, events, and shifts in the semantics of already known hashtags makes it challenging to effectively process real-time data streams, resulting in performance degradation over time.
It is interesting to note that, despite not being trend-aware, the standalone HASHET model is more robust to trend shifts than the other non-dynamic techniques. Notably, in our experiments, it is clearly outperformed by H-ADAPTS but demonstrates a degree of resilience absent in the other methods. This is due to HASHET’s ability to leverage locality and semantic affinity in the hashtag embedding space. However, when a shift introduces newly emerged hashtags, for which a latent representation is not available, or changes considerably the meaning of existing ones, making the relationships between latent vectors no longer suitable, HASHET experiences a considerable drop in performance. This leads to the widening performance gap between HASHET and H-ADAPTS shown in Figure \ref{fig:covid_weeks}.

\subsection{The 2020 US Presidential Election}

The dataset analyzed in this case study consists of $523,149$ social media posts from $183,161$ users, related to the 2020 US presidential election, which was characterized by the rivalry between candidates Joe Biden and Donald Trump.
Considered posts, published from September 1, 2020, to October 31, 2020, were filtered based on specific keywords such as ``Trump", ``Biden", and ``USElections2020". As for the COVID-19 case study, we first present the main trend shifts identified by H-ADAPTS, reported in Table \ref{tab:topics_usa}. Also in this case study, around the macro topic of the US presidential election, several micro topics emerged, related to the spread of the COVID-19 pandemic and its relationship with the presidential campaigns.

\begin{table}[h!]
\centering
\resizebox{\linewidth}{!}{
\setlength{\tabcolsep}{4pt}
\begin{tabular}{@{}lcl@{}}
\toprule
\multicolumn{1}{c}{\textbf{Start date}} & \multicolumn{1}{c}{\textbf{Topic}} & \multicolumn{1}{c}{\textbf{Top hashtags (trending set)}} \\ \midrule
\begin{tabular}[c]{@{}l@{}}Sept. 1, 2020\\(\textit{bootstrap phase})\end{tabular} & \begin{tabular}[c]{@{}c@{}} Discussion about\\ Trump's actions\\and statements\end{tabular} & \begin{tabular}[c]{@{}l@{}} \#maga, \#trump, \#covid19, \#bidenharris2020, \\\#trumpliedpeopledied, \#trump2020, \#trumpknew, \\\#trumpvirus, \#veteransforbidenharris \#werespectvets \medskip \end{tabular} \\

\begin{tabular}[c]{@{}l@{}}Sept. 30, 2020\\(\textit{first shift})\end{tabular} & \begin{tabular}[c]{@{}c@{}}Discussion about\\ the $1^{st}$ presiden-\\tial debate\end{tabular} & \begin{tabular}[c]{@{}l@{}}\#debates2020, \#presidentialdebate2020, \\\#trumpcrimefamily, \#trump, \#cashforballots,\\ \#trumptaxreturns, \#debatetuesday, \#trumpisbroke,\\  \#votehimout, \#uselections \medskip \end{tabular} \\

\begin{tabular}[c]{@{}l@{}}Oct. 4, 2020\\(\textit{second shift})\end{tabular} & \begin{tabular}[c]{@{}c@{}} Trump tested\\ positive for\\COVID-19\end{tabular} & \begin{tabular}[c]{@{}l@{}} \#covid19, \#trump, \#trumpvirus, \#covidcaughttrump, \\\#trumpcovid,  \#coronavirus, \#rosegardenmassacre,\\ \#trumphascovid, \#maga, \#vote \end{tabular} \\ \bottomrule
\end{tabular}%
\setlength{\tabcolsep}{-4pt}
}
\caption{Main trend shifts detected by H-ADAPTS in the 2020 US election case study.}
\label{tab:topics_usa}
\end{table}

During the first period, which encompasses the bootstrap window and the days before the first model adaptation, the online discussion focused on Donald Trump. Hashtags like \#maga and \#trump2020 were used to promote his reelection campaign and engage in politically-oriented discussions concerning his presidency. Among other hashtags, \#werespectvets emerged in response to allegations that Trump had privately disparaged veterans for their military service. Following this, additional hashtags like \#veteransforbidenharris and \#bidenharris2020 were increasingly used to support the Democratic candidate Joe Biden. Furthermore, as for the COVID-19 case study, hashtags like \#trumpknew, \#trumpliedpeopledied, and \#trumpvirus emerged following the revelation of Trump's conversation by the Washington Post. Notably, both case studies utilized tweets collected in September 2020, during a period when COVID-19-related issues and Trump’s policies were closely intertwined. However, the two case studies approached the topic from different perspectives, due to the distinct keywords used for data collection. Specifically, The first case study focused on the global pandemic, with Trump’s response emerging as a micro-topic, while the second centered more on election-related aspects, including Trump’s statements, actions.

The first trend shift detected by H-ADAPTS is associated with the hashtags \#debates2020, \#presidentialdebate2020, and \#debatetuesday, referring to the first presidential debate held on September 29, 2020. Many hashtags gained attention following questions and discussions that arose during the debate. Among them, the hashtag \#cashforballots was used by Trump supporters in reference to alleged electoral fraud involving offers of money in exchange for votes. Hashtags like \#votehimout and \#trumpcrimefamily reflected criticism of Trump’s presidency and related scandals. In addition, the hashtags \#trumptaxreturns and \#trumpisbroke were linked to controversies surrounding Trump’s tax returns. During the debate, he was asked about his taxes but deflected the questions, leading to criticism online.

The second shift detected by H-ADAPTS emerged when President Trump announced that he had tested positive for COVID-19 on October 2, 2020. In particular, the hashtags \#trumphascovid, \#trumpcovid, and \#covidcaughttrump were used to discuss his diagnosis, treatment, and recovery from the virus. Related to this, the hashtag \#rosegardenmassacre refers to a White House event held on September 26, 2020, where many people contracted COVID-19, including President Trump. It was used to criticize the lack of social distancing and mask-wearing at the event.

\begin{figure}[!h]
	\centering
	{\includegraphics[width=\linewidth]{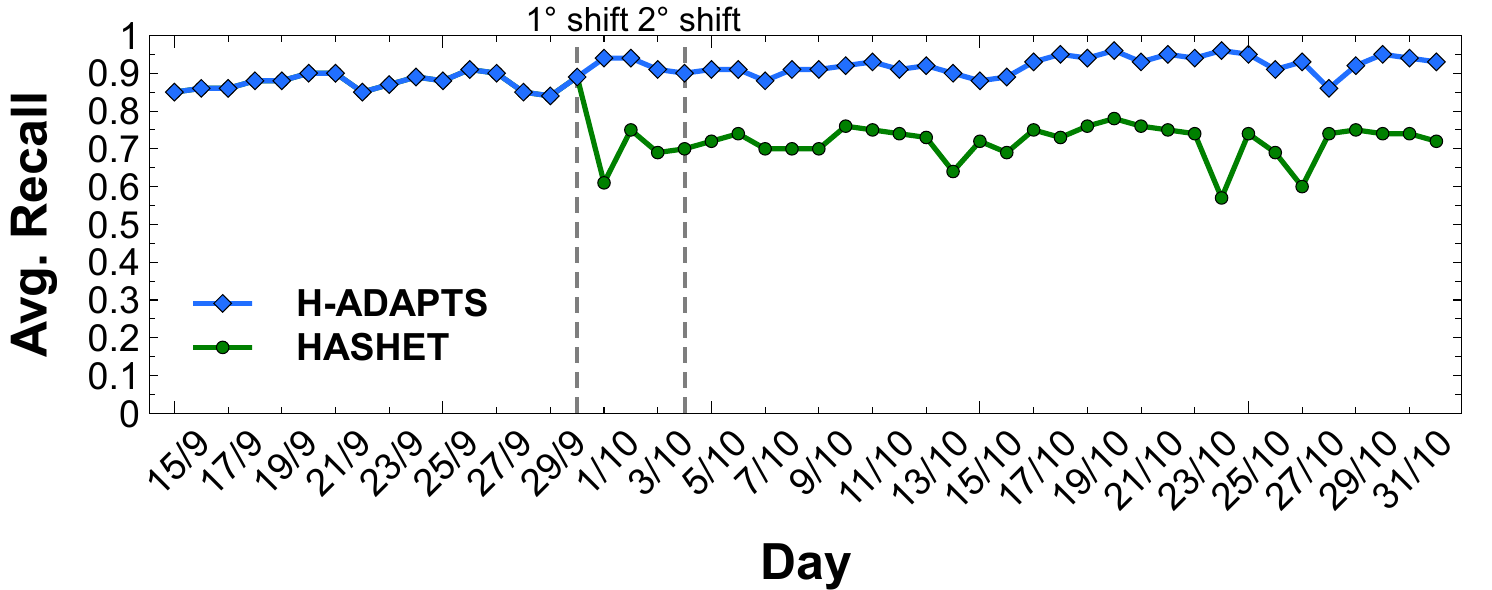}}
	\caption{Point-wise daily comparison between H-ADAPTS and HASHET for the 2020 US presidential election case study. Trend shifts are indicated by vertical dotted lines.}
	\label{fig:usa_days}	
\end{figure}

\begin{figure}[!h]
	\centering
	{\includegraphics[width=\linewidth]{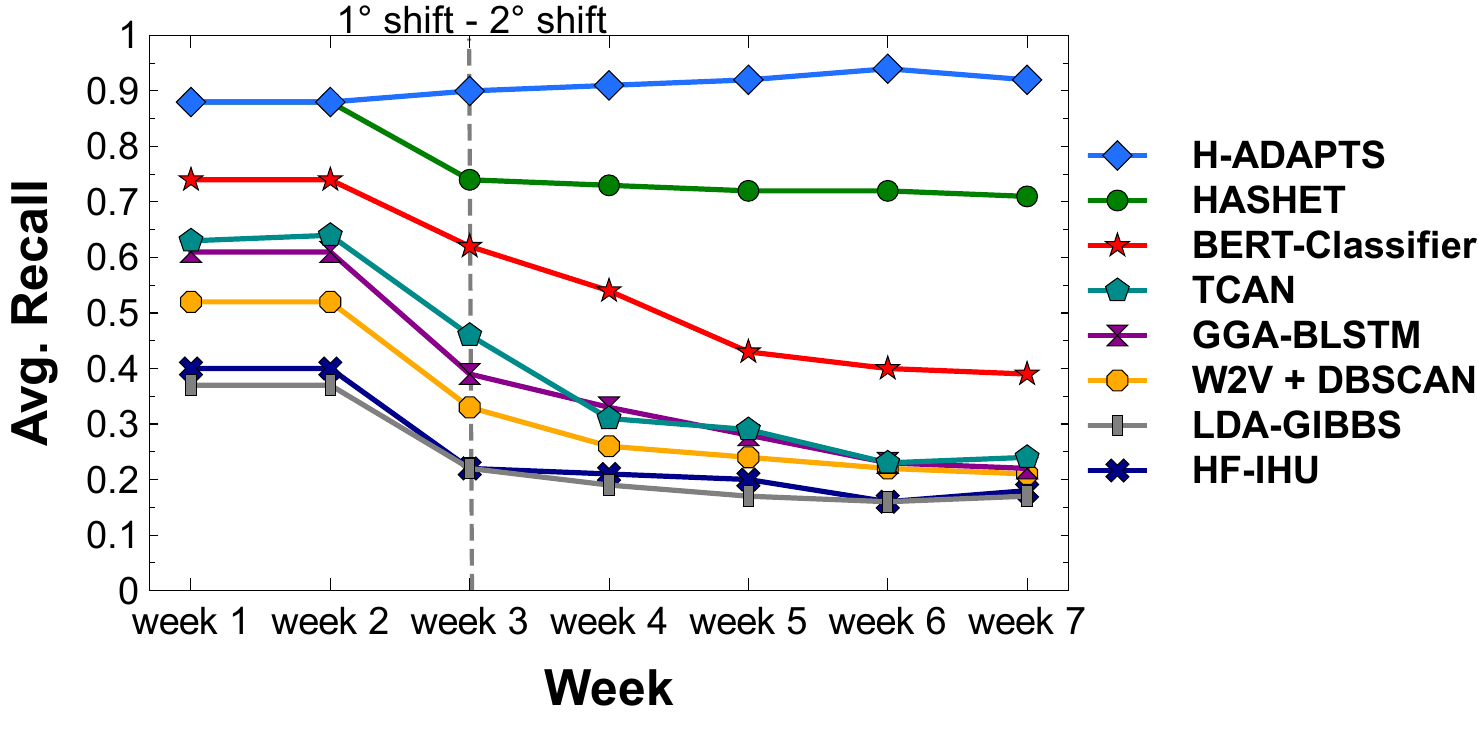}}
	\caption{Comparison with related techniques over time for the 2020 US presidential election case study, in terms of average recall. Vertical dotted lines indicate trend shifts and corresponding adaptations by H-ADAPTS.}
	\label{fig:usa_week}	
\end{figure}

By detecting and adapting to the aforementioned trend shifts, H-ADAPTS was able to recommend high-quality hashtags for all test days, as can be clearly seen in Figures \ref{fig:usa_days} and \ref{fig:usa_week}. Detected shifts, both occurring in the third week, are indicated by a vertical dotted line. On the one hand, Figure \ref{fig:usa_days} shows how the introduction of trend shift awareness leads to stable recommendation performance over time, due to the adaptation to newly emerged hashtags and topics. On the other hand, Figure \ref{fig:usa_week} shows how H-ADAPTS outperformed state-of-the-art techniques in terms of recommendation hit rate. In conclusion, the better results achieved by H-ADAPTS compared to the other techniques underpin the benefits brought by the dynamic adaptation to how social media trends emerge and evolve. This adaptiveness, enabled by the trend shift awareness of the model, is key to achieving ever-accurate recommendations, by addressing the continuous evolution of the online discussion.

\section{Conclusion}
\label{sec:conclusion}
In this work, we introduced \textit{H-ADAPTS} (\textit{Hashtag recommendAtion by Detecting and adAPting to Trend Shifts}), a dynamic hashtag recommendation methodology designed for rapidly evolving environments characterized by the continuous emergence of new trends and hashtags. H-ADAPTS extends HASHET and leverages Apache Storm to manage the high dynamism of social media conversations by detecting trend shifts in real time and adapting effectively to them. We explored various model adaptation strategies and demonstrated all trend shifts identified by H-ADAPTS through two real-world case studies, i.e., the COVID-19 pandemic and the 2020 United States presidential election. Our methodology showed robust performance even in the presence of emerging topics and hashtags, significantly outperforming state-of-the-art techniques that lack adaptive capabilities.
In future work, H-ADAPTS could be extended to other social platforms and domains, with further exploration of its integration with alternative recommendation models and the use of different detection and adaptation strategies.

\section*{Acknowledgments}
We acknowledge financial support from “National Centre for HPC, Big Data and Quantum Computing”, CN00000013 - CUP H23C22000360005, and from “PNRR MUR project PE0000013-FAIR” - CUP H23C22000860006.

\bibliography{IEEEabrv,references}

\begin{thebibliography}{10}
\providecommand{\url}[1]{#1}
\csname url@samestyle\endcsname
\providecommand{\newblock}{\relax}
\providecommand{\bibinfo}[2]{#2}
\providecommand{\BIBentrySTDinterwordspacing}{\spaceskip=0pt\relax}
\providecommand{\BIBentryALTinterwordstretchfactor}{4}
\providecommand{\BIBentryALTinterwordspacing}{\spaceskip=\fontdimen2\font plus
\BIBentryALTinterwordstretchfactor\fontdimen3\font minus \fontdimen4\font\relax}
\providecommand{\BIBforeignlanguage}[2]{{%
\expandafter\ifx\csname l@#1\endcsname\relax
\typeout{** WARNING: IEEEtran.bst: No hyphenation pattern has been}%
\typeout{** loaded for the language `#1'. Using the pattern for}%
\typeout{** the default language instead.}%
\else
\language=\csname l@#1\endcsname
\fi
#2}}
\providecommand{\BIBdecl}{\relax}
\BIBdecl

\bibitem{rangarajan2024social}
P.~K. Rangarajan, B.~M. Gurusamy, G.~Muthurasu, R.~Mohan \emph{et~al.}, ``The social media sentiment analysis framework: deep learning for sentiment analysis on social media,'' \emph{Int. J. Electr. Comput. Eng.}, 2024.

\bibitem{belcastro2020learning}
L.~Belcastro, R.~Cantini, F.~Marozzo, D.~Talia \emph{et~al.}, ``Learning political polarization on social media using neural networks,'' \emph{IEEE Access}, 2020.

\bibitem{saxena2024depth}
P.~Saxena and S.~Sharma, ``A depth examination of big data social media insights using distinctive artificial intelligence based approaches,'' in \emph{INNOCOMP}, 2024.

\bibitem{yakovlev2024recommendation}
S.~Yakovlev and N.~Shapoval, ``Recommendation of hashtags using deep learning methods based on multimodal data,'' \emph{Artificial Intelligence}, 2024.

\bibitem{HASHET-TKDD-2022}
R.~Cantini, F.~Marozzo, G.~Bruno, and P.~Trunfio, ``Learning sentence-to-hashtags semantic mapping for hashtag recommendation on microblogs,'' \emph{ACM Trans. on Knowl. Disc. Data}, 2022.

\bibitem{yuan2022recent}
L.~Yuan, H.~Li, B.~Xia, C.~Gao \emph{et~al.}, ``Recent advances in concept drift adaptation methods for deep learning.'' in \emph{IJCAI}, 2022.

\bibitem{su2024elastic}
R.~Su, H.~Guo, and W.~Wang, ``Elastic online deep learning for dynamic streaming data,'' \emph{Information Sciences}, 2024.

\bibitem{kolyszko2024dynamic}
M.~Kolyszko, M.~Buzzelli, and S.~Bianco, ``Dynamic hashtag assignment: Leveraging graph convolutional networks with class incremental learning,'' in \emph{2024 IEEE 8th Forum on Research and Technologies for Society and Industry Innovation (RTSI)}, 2024.

\bibitem{hastagger+}
B.~Shi, G.~Poghosyan, G.~Ifrim, and N.~Hurley, ``Hashtagger+: Efficient high-coverage social tagging of streaming news,'' \emph{IEEE Trans. Knowl. Data Eng.}, 2018.

\bibitem{gupta2025harnat}
D.~Gupta and S.~Chakraverty, ``Harnat-a dynamic hashtag recommendation system using news,'' \emph{Online Social Networks and Media}, 2025.

\bibitem{gupta2018unleashing}
V.~Gupta and R.~Hewett, ``Unleashing the power of hashtags in tweet analytics with distributed framework on apache storm,'' in \emph{IEEE Big Data}, 2018.

\bibitem{belcastro2022programming}
L.~Belcastro, R.~Cantini, F.~Marozzo, A.~Orsino \emph{et~al.}, ``Programming big data analysis: principles and solutions,'' \emph{J. Big Data}, 2022.

\bibitem{amen2022big}
B.~Amen, S.~Faiz, and T.-T. Do, ``Big data directed acyclic graph model for real-time covid-19 twitter stream detection,'' \emph{Pattern Recognition}, 2022.

\bibitem{chakrabarti2023hashtag}
P.~Chakrabarti, E.~Malvi, S.~Bansal, and N.~Kumar, ``Hashtag recommendation for enhancing the popularity of social media posts,'' \emph{Soc. Netw. Anal. Min.}, 2023.

\bibitem{otsuka2016hashtag}
E.~Otsuka, S.~A. Wallace, and D.~Chiu, ``A hashtag recommendation system for twitter data streams,'' \emph{Comput. Soc. Netw.}, 2016.

\bibitem{ben-lhachemi2018}
N.~Ben-Lhachemi \emph{et~al.}, ``Using tweets embeddings for hashtag recommendation in twitter,'' \emph{Procedia Comput. Sci.}, 2018.

\bibitem{godin2013topic}
F.~Godin, V.~Slavkovikj, W.~De~Neve, B.~Schrauwen \emph{et~al.}, ``Using topic models for twitter hashtag recommendation,'' in \emph{Proc. of the 23rd Int. Conf. WWW}, 2013.

\bibitem{blei2003latent}
D.~M. Blei, A.~Y. Ng, and M.~I. Jordan, ``Latent dirichlet allocation,'' \emph{J. Mach. Learn. Res.}, 2003.

\bibitem{Li2017personalized}
Y.~Li, J.~Jiang, T.~Liu, M.~Qiu \emph{et~al.}, ``Personalized microtopic recommendation on microblogs,'' \emph{ACM Trans. Intell. Syst. Technol.}, 2017.

\bibitem{li2019topical}
Y.~Li, T.~Liu, J.~Hu, and J.~Jiang, ``Topical co-attention networks for hashtag recommendation on microblogs,'' \emph{Neurocomputing}, 2019.

\bibitem{ma2018temporal}
J.~Ma, C.~Feng, G.~Shi, X.~Shi \emph{et~al.}, ``Temporal enhanced sentence-level attention model for hashtag recommendation,'' \emph{CAAI Trans. Intell. Technol.}, 2018.

\bibitem{gao2021hybrid}
J.~Gao, C.~Zhang, Y.~Xu, M.~Luo \emph{et~al.}, ``Hybrid microblog recommendation with heterogeneous features using deep neural network,'' \emph{Expert Systems with Applications}, 2021.

\bibitem{jeong2022demohash}
D.~Jeong, S.~Oh, and E.~Park, ``Demohash: Hashtag recommendation based on user demographic information,'' \emph{Expert Systems with Applications}, 2022.

\bibitem{devlin2018bert}
J.~Devlin, M.-W. Chang, K.~Lee, and K.~Toutanova, ``Bert: Pre-training of deep bidirectional transformers for language understanding,'' \emph{arXiv preprint arXiv:1810.04805}, 2018.

\bibitem{mikolov2013efficient}
T.~Mikolov, K.~Chen, G.~Corrado, and J.~Dean, ``Efficient estimation of word representations in vector space,'' \emph{arXiv preprint arXiv:1301.3781}, 2013.

\bibitem{sanh2019distilbert}
V.~Sanh, L.~Debut, J.~Chaumond, and T.~Wolf, ``Distilbert, a distilled version of bert: smaller, faster, cheaper and lighter,'' \emph{arXiv preprint arXiv:1910.01108}, 2019.

\bibitem{hinton2015distilling}
G.~Hinton, O.~Vinyals, and J.~Dean, ``Distilling the knowledge in a neural network,'' \emph{arXiv preprint arXiv:1503.02531}, 2015.

\bibitem{silalahi2022named}
S.~Silalahi, T.~Ahmad, and H.~Studiawan, ``Named entity recognition for drone forensic using bert and distilbert,'' in \emph{ICONDATA}, 2022.

\bibitem{gupta2021leveraging}
P.~Gupta, S.~Gandhi, and B.~R. Chakravarthi, ``Leveraging transfer learning techniques-bert, roberta, albert and distilbert for fake review detection,'' in \emph{Forum for Information Retrieval Evaluation}, 2021.

\bibitem{sharma2021recent}
G.~Sharma, V.~Tripathi, and A.~Srivastava, ``Recent trends in big data ingestion tools: A study,'' in \emph{Proc. of RICE 2020}, 2021.

\bibitem{Elbedwehy2022}
S.~Elbedwehy, C.~Thron, M.~Alrahmawy, and T.~Hamza, \emph{Real-Time Detection of First Stories in Twitter Using a FastText Model}, 2022.

\bibitem{7530084}
S.~Chintapalli, D.~Dagit, B.~Evans, R.~Farivar \emph{et~al.}, ``Benchmarking streaming computation engines: Storm, flink and spark streaming,'' in \emph{IEEE IPDPS Workshops}, 2016.

\bibitem{8509390}
J.~Karimov, T.~Rabl, A.~Katsifodimos, R.~Samarev \emph{et~al.}, ``Benchmarking distributed stream data processing systems,'' in \emph{IEEE ICDE}, 2018.

\bibitem{7841533}
M.~A. Lopez, A.~G.~P. Lobato, and O.~C. M.~B. Duarte, ``A performance comparison of open-source stream processing platforms,'' in \emph{IEEE GLOBECOM}, 2016.

\bibitem{mahajan2018exploring}
D.~Mahajan, R.~Girshick, V.~Ramanathan, K.~He \emph{et~al.}, ``Exploring the limits of weakly supervised pretraining,'' in \emph{Proc. of ECCV}, 2018.

\end{thebibliography}

\bibliographystyle{IEEEtran}

\newpage

\vskip -20pt plus -1fil
\begin{IEEEbiography}[{\includegraphics[width=1in,height=1.25in,clip,keepaspectratio]{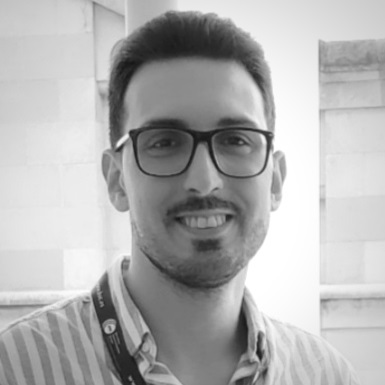}}]{Riccardo Cantini} is a computer engineering researcher at the University of Calabria, Italy. 
In 2021-2022 he was a visiting researcher at the Barcelona Supercomputing Center (BSC).
His research interests include deep learning, with a focus on Large Language Models and sustainable AI, and big social data analysis, targeting politically polarized data and the efficient execution of data-intensive applications in high-performance distributed environments.
\end{IEEEbiography}
\vskip -20pt plus -1fil
\begin{IEEEbiography}[{\includegraphics[width=1in,height=1.25in,clip,keepaspectratio]{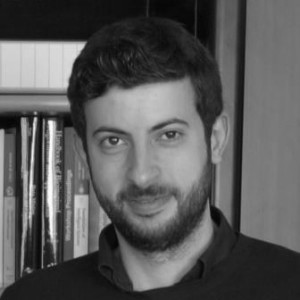}}]{Fabrizio Marozzo} is an associate professor of computer engineering at the University of Calabria. 
In 2011-2012 he visited the Barcelona Supercomputing Center for a research internship in the Computer Sciences department. 
He is a member of the editorial boards of several journals including IEEE Access, IEEE Transactions on Big Data, and Journal of Big Data. His research focuses on big data, distributed computing, cloud and edge computing, and social media analysis. He is a Senior Member of IEEE.
\end{IEEEbiography}
\vskip -20pt plus -1fil
\begin{IEEEbiography}[{\includegraphics[width=1in,height=1.25in,clip,keepaspectratio]{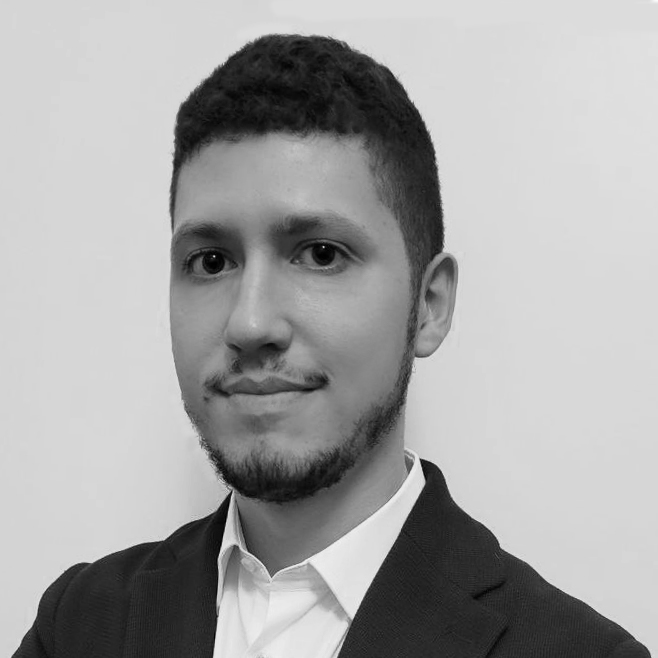}}]{Alessio Orsino} is a research fellow at the University of Calabria, Italy, where he obtained his Ph.D. in Information and Communication Technologies in 2025. In 2023, he was a visiting researcher at the Department of Computer Science and Technology at the University of Cambridge. His research interests include big data analysis, green and sustainable AI, edge computing, and high-performance computing.
\end{IEEEbiography}
\vskip -20pt plus -1fil
\begin{IEEEbiography}[{\includegraphics[width=1in,height=1.25in,clip,keepaspectratio]{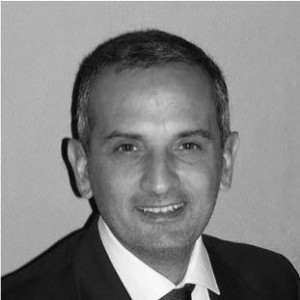}}]{Domenico Talia} is a full professor of computer engineering at the University of Calabria and a honorary professor at Noida University. He is a member of the editorial boards of Future Generation Computer Systems, IEEE Transactions on Parallel and Distributed Systems, and ACM Computing Surveys. His research interests include parallel and distributed data mining, cloud computing, machine learning, big data, peer-to-peer systems, and parallel programming models. He is a Senior Member of IEEE and ACM.
\end{IEEEbiography}
\vskip -20pt plus -1fil
\begin{IEEEbiography}[{\includegraphics[width=1in,height=1.25in,clip,keepaspectratio]{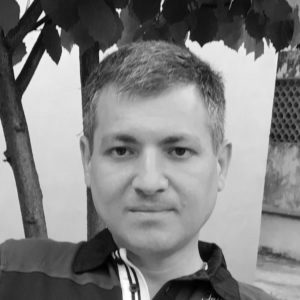}}]{Paolo Trunfio} is a full professor of computer engineering at the University of Calabria. In 2007 he was a visiting researcher at the Swedish Institute of Computer Science. 
He is in the editorial board of IEEE Transactions on Cloud Computing, ACM Computing Surveys, Future Generation Computer Systems, and Journal of Big Data. His research interests include cloud computing, social media analysis, distributed knowledge discovery, and peer-to-peer systems. He is a Senior Member of IEEE and ACM. 
\end{IEEEbiography}

\vfill

\end{document}